\documentclass[twocolappendix]{emulateapj}

\usepackage[usenames,dvipsnames]{color}
\usepackage{soul,color,amsmath,comment}
\newcommand{\hilight}[1]{{#1}}

\shorttitle{A 1D Rayleigh-Taylor Model}
\shortauthors{Duffell}

\begin{document}

\title{A one-dimensional model for Rayleigh-Taylor Instability in Supernova Remnants}

\author{Paul C. Duffell}
\affil{Astronomy Department and Theoretical Astrophysics Center, University of California, Berkeley}
\email{duffell@berkeley.edu}

\begin{abstract}

This study presents a method for approximating the multidimensional effects of Rayleigh-Taylor instability as a modification of the one-dimensional hydro equations.  This modification is similar to the Shakura-Sunyaev $\alpha$ prescription for modeling the coarse-grained effects of turbulence in astrophysical disks.  The model introduces several dimensionless tunable parameters that are calibrated by comparing with high-resolution two-dimensional axisymmetric numerical calculations of Rayleigh-Taylor unstable flows.  A complete description of the model is presented, along with a handful of test problems that demonstrate the extent to which the one-dimensional model is able to reproduce multidimensional effects.

\end{abstract}


\section{Introduction}
\label{sec:intro}

Many astrophysical outflows are Rayleigh-Taylor (RT) unstable \citep{1978ApJ...219..994C}.  This instability affects the properties of the flow, significantly enough that one-dimensional (1D) calculations assuming spherical symmetry may not match up with direct observations of supernova ejecta.  RT causes disruption of sharp density jumps at contact discontinuities, mixing of ejecta with the circumstellar medium (CSM), interactions between the unstable region and the reverse shock, and line broadening due to turbulent fluctuations.  This turbulence may also generate magnetic fields via small-scale turbulent dynamo \citep{1995ApJ...453..332J, 2013ApJ...775...87D}, and might even alter the forward shock propagation, if there is significant cooling in the shock front \citep{2001ApJ...560..244B, 2014ApJ...791L...1D}.

Because of the importance of these multidimensional effects, many numerical investigations have been launched over the past several decades to determine the properties of RT-unstable flows.  The very first two-dimensional (2D) numerical calculations were by \cite{1978ApJ...219..994C}, who studied structure formation in the context of Type II supernovae.  Higher-resolution 2D results became possible decades later, and the problem was more accurately tackled by \cite{1992ApJ...392..118C}, who calculated growth rates both analytically and numerically.  \cite{1995ApJ...453..332J} studied two and three-dimensional (3D) RT with magnetic fields in an idealized context, finding that magnetic fields affected the growth rate and that RT easily amplifies magnetic fields.  \cite{1996ApJ...465..800J} demonstrated in the supernova context (in 2D) that RT could cause these fields to align with turbulent structures, affecting polarization of synchrotron emission.  \cite{2000ApJ...528..989K} studied the difference between 2D and 3D, but still in a local sense (looking at single-mode perturbations).  That study found that the growth of RT is $30-35\%$ stronger in 3D than in 2D.

\cite{2001ApJ...560..244B} first demonstrated the importance of cooling on the dynamics of RT.  Specifically, cosmic rays provide significant cooling in shocks, resulting in shallower pressure gradients, allowing the Rayleigh-Taylor fingers to catch up to the forward shock.  This was demonstrated first by varying the adiabatic index, showing a dramatic change in the dynamics.  The importance of cosmic rays on the dynamics has also been shown observationally \citep{2005AAS...20717212W}.  \cite{2010AnA...515A.104F} and \cite{2010AnA...509L..10F} performed the first 3D global studies of RT, that also included a prescribed model for cosmic ray cooling (rather than varying the adiabatic index).  RT has also been studied in the relativistic case, in the context of gamma ray bursts \citep{2013ApJ...775...87D, 2014ApJ...791L...1D}.

On the experimental side, RT has been widely explored by various groups \citep[e.g.][]{1996PhRvL..76.4536B, 1997PhPl....4.1994R, 1999Sci...284.1488R, 2000PhFl...12..304D, 2001PhPl....8.2446R, 2005PhRvL..95u5001S, 2010PhPl...17e2709K, 2012PhPl...19h2708C}.  This is partly due to its importance in inertial confinement fusion \citep{1975PhRvL..34.1273L, 1982PhFl...25.1653V, 1995PhPl....2.3933L}.  Richtmyer-Meshkov instability is also important in this context \citep{2003PhPl...10.1931G, 2011PhFl...23i5107T}.

\hilight{Given RT's importance, and given that many numerical studies of supernova dynamics are still carried out in 1D, one would expect the development of a 1D model that could be used by theorists numerically studying supernovae using 1D radiation hydrodynamics codes.  Surprisingly, the only method which could be easily introduced into a standard 1D hydro code was developed over 40 years ago \citep{1973MNRAS.161...47G}, before multidimensional numerical studies of RT were possible.}

\hilight{There have been numerous studies which attempt to theoretically model the nonlinear effects of RT in various ways \citep[e.g.][]{1989PhRvA..39.5812H, 1994PhRvL..72.2867A, 2004PhFl...16.1668D, 1995PhPl....2.2465S, 1996PhPl....3.3073O, 2001PhPl....8.2883O}, using both experimental and numerical results.  More general turbulence models have been suggested, e.g. the ``One-Dimensional Turbulence" formulation of \cite{1999JFM...392..277K} \citep[see also][]{2001PhFl...13..702W, ashurst2005one}, which uses a Monte Carlo approach in an attempt to model detailed statistical properties of the turbulence, including a Kolmogorov $5/3$ cascade.  Most of these models generate a series of ordinary differential equations (ODEs) describing the evolution of various properties of a mixing layer.  In contrast, Gull's model was expressible as a set of partial differential equations (PDEs) (adding additional terms to Euler's equations), which is useful for incorporation into existing 1D hydro codes.  However, only a handful of PDE turbulence models have been built specifically for RT.}

\hilight{A few notable such RT models include \cite{youngs1989modelling}, a multiphase model for experiments using multiple immiscible fluids, and the ODE formulation of \cite{ramshaw1998simple}, which has also been cast in PDE form \citep{ramshaw2000implementation}.  Both of these models have two drawbacks which make it difficult to apply them to the supernova context.  First, both model the system using a multiphase approach, suggesting an initial condition of two separate fluids, and hence foreknowledge of the position of the unstable layer.  Ideally, an RT model should allow unstable layers to grow anywhere they happen to emerge, treating every fluid element in the system equally.  Such an approach requires a single-fluid model.  Secondly, an expanding outflow sets limits on the maximal allowed coherent length scale in the turbulence, and this does not appear to be accounted for in either of the models of \cite{youngs1989modelling} or \cite{ramshaw2000implementation}.  Kerstein's ODT model was also applied to both RT and Richtmeyer-Meshkov instability \citep{gonzalez2013reactive, jozefik2015towards}, but this model has similar drawbacks as the models of Youngs and Ramshaw.  Furthermore, the version applied to RT requires specific knowledge of the position of the mixing layer, and tracks its growth with time as part of the model; it is not clear how one would extend this to a model where multiple mixing layers could spontaneously form anywhere that instability is triggered.  Moreover, the full reproduction of all details of the statistics of the turbulence may be overkill for the purposes of a 1D supernova remnant calculation.}

\hilight{In practice, when evolving supernovae in 1D, one is often forced to artificially ``smear out" regions which are RT unstable in post-processing \citep[e.g.][]{1988ApJ...329..820P, 2009ApJ...703.2205K, 2010ApJ...724..341H}.  Such an operation is obviously undesirable, but it may be necessary if nothing else is done to model the effects of RT.}

\hilight{Building on Gull's 1973 method, an improved 1D model is presented in this work, one that is informed and calibrated by multidimensional numerical RT calculations.  Following Gull's original idea, an additional scalar quantity is evolved along with the usual hydro variables.  This field, $\kappa$, represents the magnitude of turbulent RT fluctuations, analogous to the Shakura-Sunyaev $\alpha$ prescription in disk physics \citep{1973AnA....24..337S}.  This is also a similar approach to that employed to deal with convective regions in 1D stellar evolution codes, and turbulence-induced mixing in type Ia supernovae \citep{2009ApJ...704..255W, 2011ApJ...734...37W}.}

$\kappa$ is prescribed a growth rate in RT-unstable regions of the flow, and is used to calculate a local diffusion constant, $\eta$, that causes mixing of all conserved variables.  This surprisingly simple prescription is enough to produce dynamics reasonably consistent with true multidimensional RT calculations, sufficiently so that the model could be considered an improvement to a purely 1D calculation.

This one-dimensional model is detailed in Section \ref{sec:model}, and is calibrated and tested in several supernova contexts in Section \ref{sec:tests}.  A summary is presented in Section \ref{sec:summary}.

\section{Model Description}
\label{sec:model}

This study will produce an augmented system of hydro equations, in an attempt to reproduce some of the coarse-grained effects in a one-dimensional model.  The unmodified spherical 1D hydro equations (in conservation-law form) are

\begin{equation}
\partial_t {( \rho )} + r^{-2} ( r^2 ( \rho v ) )' = 0
\label{eqn:bare1}
\end{equation}
\begin{equation}
\partial_t {( \rho v )} + r^{-2} ( r^2 ( \rho v^2 + P ) )' = 2 P / r
\label{eqn:bare2}
\end{equation}
\begin{equation}
\partial_t {( \epsilon_{tot} )} + r^{-2} ( r^2 ( \epsilon_{tot} + P ) v )' = 0,
\label{eqn:bare3}
\end{equation}

where primes denote radial derivatives, $\rho$ is density, $v$ is the radial velocity, $P$ is pressure, and $\epsilon_{tot}$ is the total energy density, defined as:

\begin{equation}
\epsilon_{tot} = \frac12 \rho v^2 + \epsilon_{Th}
\end{equation}

where $\epsilon_{Th}$ is the thermal energy density, and the equations are closed by an equation of state:

\begin{equation}
P = (\gamma - 1) \epsilon_{Th},
\end{equation}

where $\gamma$ is the adiabatic index ($\gamma = 5/3$ here).  Additionally, one can evolve passive scalars (e.g. representing various nuclear abundances) as an additional conservation law:

\begin{equation}
\partial_t {( \rho X )} + r^{-2} ( r^2 ( \rho X v ) )' = 0.
\end{equation}

In this study, a passive scalar X will be used to mark the separation between ejecta and the CSM.  X=1 for the ejecta and X=0 for the CSM.  This will be used to measure mixing between the two fluids.\\

\begin{figure}
\epsscale{1.27}
\plotone{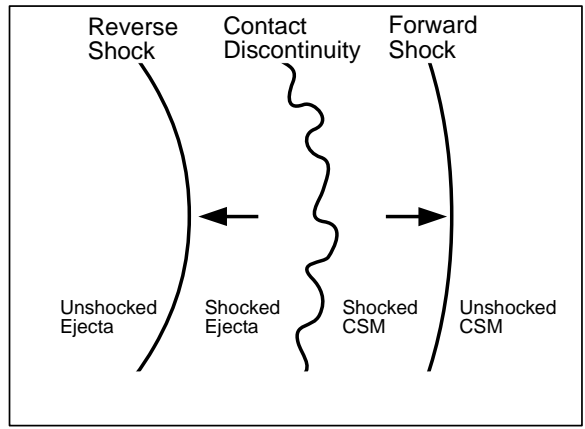}
\caption{ Cartoon depiction of the unstable contact discontinuity between ejecta and CSM in the outflow from a supernova.
\label{fig:cartoon} }
\end{figure}

\begin{figure*}
\epsscale{1.0}
\plotone{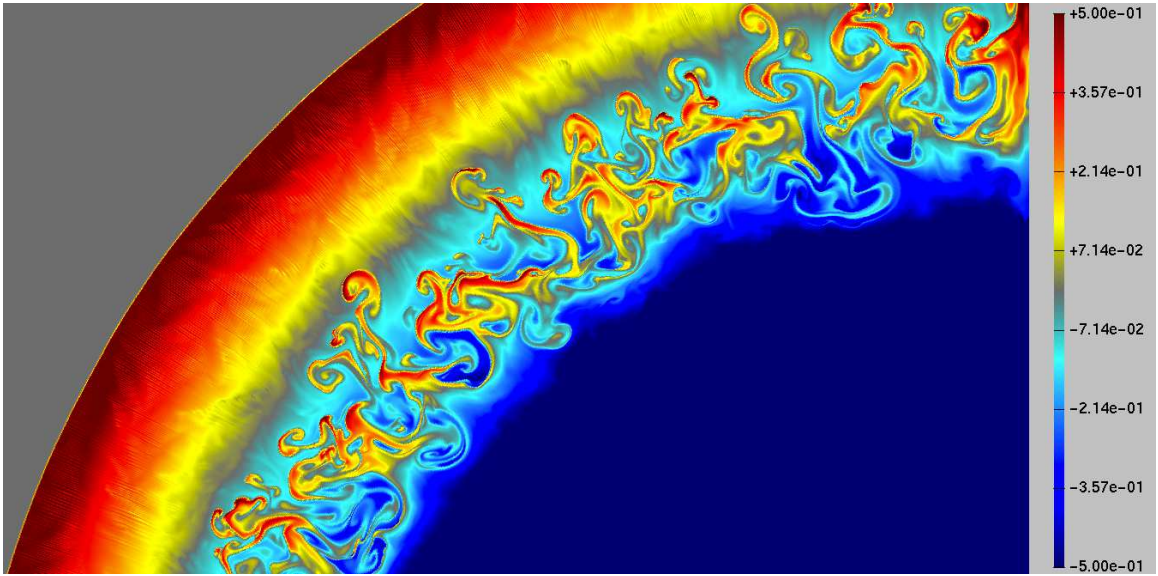}
\caption{ RT turbulence in a calculation of the idealized supernova at time $t = 1$ (Test \#1, Section \ref{sec:test1}; color denotes log of density).
\label{fig:idealsn1} }
\end{figure*}

The RT instability occurs at large density gradients, typically at contact discontinuities between two fluids.  In the supernova context, the two fluids are typically ejecta (mass ejected in the explosion) and the CSM (external mass swept up by the outflow).  As the ejecta sweeps up the CSM, shocks are generated by the collision.  A forward shock rushes ahead into the CSM, and a reverse shock propagates back into the ejecta, notifying fluid elements in the ejecta that it is time to decelerate \citep{1974ApJ...188..335M, 1982ApJ...258..790C}.  In the hot region between the two shocks resides the contact discontinuity, separating ejecta from CSM (Figure \ref{fig:cartoon}).  For typical outflows at early times, this contact discontinuity is unstable (see also Figure \ref{fig:idealsn1}).

The instability is driven by the effective gravitational force felt in the decelerating reference frame of the contact discontinuity.  In practical terms, unstable configurations exist whenever the pressure gradient has an opposite sign to the density gradient \citep{1961hhs..book.....C}.  In this case, the pressure gradient is positive, as the ejecta is decelerating, and at early times the ejecta is more dense than the CSM, so the density gradient is sharply negative across the contact discontinuity, at least until a significant amount of the CSM is swept up.

In order to account for these multidimensional effects, an additional scalar $\kappa$ is evolved with the 1D flow.  $\kappa$ represents the relative strength of turbulent fluctuations, similar to the Shakura-Sunyaev $\alpha$ viscosity.

The scalar quantity $\kappa$ will be interpreted as

\begin{equation}
\kappa = \langle \delta u^2 \rangle / c_s^2,
\label{eqn:kappa}
\end{equation}

where $\delta u$ represents turbulent fluctuations of velocity about the background mean (averaged over a sufficiently large patch), and $c_s$ is the local sound speed, $c_s^2 = \gamma P/\rho$.  Note that $\kappa$ is of order the ratio of kinetic to thermal energy in the turbulence:

\begin{equation}
\kappa = \rho \delta u^2 / (\gamma P) = {2 \over \gamma} {\frac12 \rho \delta u^2 \over P} = {2 \over \gamma (\gamma-1)} {\epsilon_K \over \epsilon_{Th}}
\label{eqn:kappa0}
\end{equation}

In stable regions of the flow, $\kappa$ will be evolved as a passive scalar:

\begin{equation}
\partial_t {( \rho \kappa )} + r^{-2} ( r^2 ( \rho \kappa v ) )' = 0.
\label{eqn:kappa1}
\end{equation}

Note that this implicitly assumes that $\kappa$ stays fixed, even as a fluid element expands or contracts.  This assumption is based on the idea that random kinetic fluctuations should have an effective adiabatic index of $5/3$, and should therefore maintain a constant proportion with the thermal energy.

This also neglects many source terms which would appear in a formal Reynolds decomposition of the fluid equations.  Viscous dissipation is neglected, as this entire treatment assumes zero viscosity, due to the large Reynolds numbers of these systems.  Shear production and Buoyant production of turbulence are also neglected here, and these terms should be included in a more complete model, which would more accurately describe the evolution of the turbulence.  In the present study, it is assumed that Rayleigh-Taylor instability dominates the growth of turbulence.

Equation (\ref{eqn:kappa1}) also does not take into account decay due to turbulent dissipation.  This can potentially be added as a source term, as will be described later.  It bears mentioning, however, that the present work is restricted to 2D numerical calculations for calibration of this model, and therefore turbulent decay will not be accurately captured in this study.  \cite{1999ApJ...524..169M}  showed that turbulence damps efficiently, and therefore this is an important consideration.

The goal is to augment the 1D equations (\ref{eqn:bare1}-\ref{eqn:bare3}) with additonal terms which mimic the effects of turbulence.  Although the turbulence is subsonic, it propagates through steep density gradients, so that there are not just velocity perturbations $\delta \vec u$, but also density fluctuations $\delta \rho$.  These fluctuations are significant enough that they are the dominant source of diffusion of all quantities (as opposed to the standard Reynolds stress terms $\rho \left<\delta u_i \delta u_j \right> \ll P$).

For example, consider the equation of mass continuity,

\begin{equation}
\partial_t \rho + \vec \nabla \cdot ( \rho \vec v ) = 0.
\end{equation}

the mass flux $\rho \vec v$ can be rewritten in terms of mean and fluctuating quantities:

\begin{equation}
F_{\rho} = \left< \rho \vec v \right> = \left< \rho_{\rm avg} \vec v_{\rm avg} + \delta \rho \vec v_{\rm avg} + \rho_{\rm avg} \delta \vec u + \delta \rho \delta \vec u \right>
\end{equation}

where the subscript ``avg" denotes averaged quantities (this subscript will now be dropped for simplicity).  In the overall average, the first term is the bulk mass flux, and the second and third terms average to zero.  The last term 

\begin{equation}
\delta F_{\rho} = \left< \delta \rho \delta \vec u \right>
\end{equation}

is nonzero if density fluctuations are correlated with velocity fluctuations.  They should indeed correlate, if there is a density gradient in the flow.  Assume that the fluctuation in density is given by the density that the fluid element would have carried one correlation time ago:

\begin{equation}
\delta \rho \sim - \tau \delta \vec u \cdot \vec \nabla \rho,
\label{eqn:corr}
\end{equation}

where $\tau$ is a correlation time.  Given this assumption, mass flux can be re-expressed as

\begin{equation}
F_{\rho} = \rho v - \tau \left< \delta u^2 \right> \rho' = \rho v - \eta \rho'.
\label{eqn:etadef}
\end{equation}

where $\eta = \tau \left< \delta u^2 \right>$ is a diffusion constant.  Similar arguments can be made to build diffusive fluxes for all conserved quantities\footnote{It should be noted that the precise form of these terms depends on assumptions made on how the conserved densities change as they are carried along by a fluid element.  In this study, such effects were neglected, but future improvements should include these effects; for example, as a subsonic fluid element propagates through a pressure gradient, the density adjusts to maintain specific entropy.  Therefore, a more accurate form for (\ref{eqn:corr}) should be $- \tau P^{1/\gamma} \delta \vec u \cdot \nabla (\rho P^{-1/\gamma})$ }.  To summarize, neglecting growth or decay of turbulence, the equations should now have the form

\begin{equation}
\partial_t {( \rho )} + r^{-2} ( r^2 ( \rho v - \eta \rho' ) )' = 0
\label{eqn:diff1}
\end{equation}
\begin{equation}
\partial_t {( \rho v )} + r^{-2} ( r^2 ( \rho v^2 + P - \eta (\rho v)' ) )' = 2 P / r
\label{eqn:diff2}
\end{equation}
\begin{equation}
\partial_t {( \epsilon_{\rm tot} )} + r^{-2} ( r^2 ( (\epsilon_{\rm tot} + P) v - \eta \epsilon_{\rm tot}' ) )' = 0,
\label{eqn:diff3}
\end{equation}
\begin{equation}
\partial_t {( \rho X)} + r^{-2} ( r^2 ( \rho X v - \eta (\rho X)' ) )' = 0
\label{eqn:diff4}
\end{equation}
\begin{equation}
\partial_t {( \rho \kappa)} + r^{-2} ( r^2 ( \rho \kappa v - \eta (\rho \kappa)' ) )' = 0
\label{eqn:diff5}
\end{equation}

where $\eta = \tau \left< \delta \vec u^2 \right> = \lambda \left< | \delta \vec u | \right> = \lambda \sqrt{\kappa} c_s$, where $\lambda$ is a correlation length, discussed below.

Growth of $\kappa$ is assumed to only take place in unstable regions.  Instability occurs wherever the gradients of pressure and density have opposite signs.  It should be noted that $\kappa$ is only meant to represent the fluctuations of the largest-scale modes, as these modes dominate the mixing and turbulent stresses.  The length scale $\lambda$ of these largest modes can be estimated by the following exercise:

In order to have coherent structures in an expanding flow, the flow must not be expanding too quickly for the turbulent velocities to merge structures together on a given scale.  The largest-scale modes will be at the critical scale such that the relative expansion velocity between two fluid elements separated by $\lambda$ in the background flow is of order the turbulent velocities:

\begin{equation}
\lambda \nabla \cdot v \sim \delta u
\end{equation}

Assuming the background flow is expanding radially at a velocity of order the sound speed, and velocity fluctuations are given by (\ref{eqn:kappa}), 

\begin{equation}
\nabla \cdot v \sim c_s/r, ~~\delta u \sim \sqrt{\kappa} ~ c_s.
\end{equation}

This provides an estimate for the largest coherent scale of the turbulence:

\begin{equation}
\lambda \sim \sqrt{\kappa} ~ r.
\end{equation}

\hilight{Note that this estimate may be inaccurate for a number of reasons (for example, the incorrect assumption that the flow expands at the sound speed, when actually there can be a large disparity between sound speeds across the contact discontinuity \citep{1999ApJS..120..299T, 2000ApJS..128..403T}).  In reality, any estimate of $\lambda$ from local quantities is probably flawed, as $\lambda$ can cover large scales, and is not necessarily locally calculable.  Previous 1D RT models \citep[e.g.][]{youngs1989modelling, ramshaw2000implementation, gonzalez2013reactive} were designed in an idealized ``infinite slab" context, where arbitrarily large wavelengths are allowed.  Here, the spherical geometry and the expanding flow set the scale $\lambda$.}

\hilight{The nonlinear growth rate of Rayleigh-Taylor can be determined by imagining a mixing layer with density jump between two fluids with density $\rho_1$ and $\rho_2$.  Under constant acceleration $g$, the width of the mixing layer, denoted by $h$, grows quadratically with time in the nonlinear regime \citep{dimonte2004comparative}:}

\begin{equation}
h(t) = \alpha g \mathcal{A} t^2
\label{eqn:hnonl}
\end{equation}

\hilight{where $\mathcal{A} = (\rho_1 - \rho_2)/(\rho_1 + \rho_2)$ is the Atwood number, $g$ is the acceleration felt by the mixing layer, and $\alpha$ is a constant, empirically determined to have a value between $0.03$ and $0.08$.  The growth rate of this mixing layer is then}

\begin{equation}
\Gamma_{\rm RT} = \dot h / h = 2/t = 2 \sqrt{ \alpha g \mathcal{A} / h }.
\end{equation}

\hilight{Assume that the density varies linearly from $\rho_1$ to $\rho_2$ over the width $h$.  Given that $\mathcal{A} / h$ = $\Delta \rho / ( 2 \rho h )$, (where $\Delta \rho$ is the density jump and $\rho = (\rho_1 + \rho_2)/2$ is the average density in the mixing layer), and $\Delta \rho / h = -\rho'$, the derivative of density in the mixing layer, and $g = P'/\rho$, this reduces to a growth rate of}

\begin{equation}
\Gamma_{\rm RT} \sim \sqrt{ - \rho' P' / \rho^2 }.
\label{eqn:grow2}
\end{equation}

\hilight{where the constants have been dropped, and the minus sign is necessary as RT only grows when $\rho'$ and $P'$ have opposite signs.  This growth rate translates into a source term for Equation (\ref{eqn:kappa1}):}

\begin{equation}
S^{+}_{\kappa} = (A+B\kappa) \Gamma_{RT} \rho = (A+B\kappa) \sqrt{ - P' \rho' },
\end{equation}

where $A$ and $B$ are arbitrary dimensionless constants, and the source term is only added in regions where the pressure and density gradients have opposite signs, i.e. in regions where the square root evaluates to a real number.  Note that this gives exponential growth for sufficiently large $\kappa > A/B$, but the small-amplitude growth phase is linear.  This is because the goal is to model the nonlinear growth phase of the large-scale turbulence in supernova remnants, after a significant amount of CSM mass has been swept up by the ejecta.  In numerical studies it has been found that the saturated instability is independent of the seed perturbation \citep{2013ApJ...775...87D}, and therefore small-scale turbulence amounts to some nonzero growth rate at early times, after which exponential growth takes over.  This might not apply at early times in the evolution of the supernova.

Turbulent decay can be taken to be exponential, with a decay timescale proportional to an eddy turn-over:

\begin{equation}
\tau_{decay} \sim \lambda / |\delta u|.
\end{equation}

where $\lambda$ is the characteristic eddy size.  Therefore, a sink term $S_{\kappa}^{-}$ should also be added:

\begin{equation}
S^{-}_{\kappa} = - D \rho \kappa / \tau_{decay} = - D \rho \kappa c_s / r.
\end{equation}

in the current study, this term is turned off, $D=0$, as the results are only compared with 2D turbulence, which should not exhibit significant decay.  The term is included in the equations, however, so that future 3D studies can provide a decay constant.  Finally, the value of $\kappa$ is used to calculate a kinematic diffusion constant $\eta$.  This diffusion constant appears in equation (\ref{eqn:etadef}) as $\eta \sim \tau \left< \delta u^2 \right>$, or 

\begin{equation}
\eta \sim |\delta u| \lambda.
\end{equation}

(this is also clear from dimensional grounds).  Given $\lambda$ and using $|\delta u| = \sqrt{\kappa} c_s$,

\begin{equation}
\eta = C ( \kappa c_s r ),
\end{equation}

where $C$ is another dimensionless constant that will be calibrated using multidimensional calculations.

All conserved quantities (including $\rho \kappa$) are mixed according to the diffusion constant $\eta$.  To summarize, the complete augmented 1D system of equations can be expressed in conservation-law form as

\begin{equation}
\partial_t U + r^{-2} ( r^2 ( F - \eta U' ) )' = S,
\end{equation}

giving one equation for each conserved variable $U$:

\begin{equation}
U = \left\{ \rho , \rho v , \epsilon_{tot}, \rho X , \rho \kappa \right\}.
\end{equation}

The fluxes are given by

\begin{equation}
F = \left\{ \rho v , \rho v^2 + P , (\epsilon_{tot} + P) v, \rho X v , \rho \kappa v \right\}
\end{equation}

and the source terms are

\begin{equation}
S = \left\{ 0 , 2 P / r , 0 , 0 , S^{+}_{\kappa} + S^{-}_{\kappa} \right\},
\end{equation}

where

\begin{equation}
S^{+}_{\kappa} = (A+B\kappa) \sqrt{ \text{max}(- P' \rho',0) },
\end{equation}

\begin{equation}
S^{-}_{\kappa} = - D \rho \kappa c_s / r,
\end{equation}

\begin{equation}
\eta = C \kappa c_s r.
\end{equation}

The choice of $\lambda$ is nontrivial, as it is an inherently global property of the flow, and therefore cannot be computed from local quantities.  \cite{1973MNRAS.161...47G} effectively made the choice that $\lambda$ be associated with the zone size $\lambda \sim \Delta r$, but this is of course problematic, as the solution might then depend on resolution.

In this study, the assumption is made that $\lambda = \sqrt{\kappa} r$, an assumption that is reasonably well-motivated based on the arguments above, but an assumption nonetheless.  On the other hand, reasonable results are still possible for appropriate choices of the constants $A$, $B$, $C$, and $D$.  A more accurate model may be possible using a better-informed prescription for $\lambda$.  For the choice $\lambda = \sqrt{\kappa} r$, two-dimensional RT is most accurately approximated by the constants $A = 1.18 \times 10^{-5}$, $B = 1.2$, $C = 0.102$, $D=0$, as will be demonstrated in the next section.\\

Implementation in an existing 1D hydro code is very straightforward, especially if the code explicitly evolves the conservative form of the equations.  For a given conserved quantity, $U$, the flux of $U$ is simply modified by the replacement

\begin{equation}
F(U) \rightarrow F(U) - \eta U'.
\end{equation}

All other aspects of the code (e.g. Riemann solvers) remain unchanged.  This way of adding the diffusive term is explicit, which would be problematic for large diffusion constants, but as $\kappa$ is typically small $\sim 1-10\%$, this does not pose a major problem.   The only additional issue is a correction to the timestep; in addition to the Courant condition, one must enforce the criterion

\begin{equation}
\Delta t < \Delta r^2 / \eta.
\end{equation}

Again, for small $\eta$ this is not a major problem.  Such timestep restrictions might also be avoided using an implicit time integrator.

\section{Numerical Tests}
\label{sec:tests}

\hilight{The numerical tests in this study are all performed in a similar way.  Three calculations are performed for each test: a 1D calculation with no RT model, a 2D axisymmetric calculation with real turbulence, and a 1D calculation with the RT model specified above.  These tests are used for both calibration and evaluation of the method.  Upon performing the first test, the following values for the tunable constants are found: $A = 1.18 \times 10^{-5}$, $B = 1.2$, $C = 0.102$, $D=0$.  These chosen values for the constants are used for all three numerical tests.  It should be noted that the dynamics were found to be largely insensitive to the value of $A$, so long as it is small but nonzero.  This could be linked to the fact that the large-scale dynamics of RT have been noted to be largely insensitive to initial seed perturbations \citep{2013ApJ...775...87D}.  However, this statement may depend on the wavelength of the initial seed perturbation \citep{dimonte2004comparative}}

The 2D axisymmetric calculations are performed using the highly accurate JET code \citep{2011ApJS..197...15D, 2013ApJ...775...87D}.  JET is a moving-mesh hydrodynamics code which is specifically tailored to the study of astrophysical jets and outflows.  The numerical method is based on the relativistic moving-mesh TESS code \citep{2011ApJS..197...15D}, but with a more restricted, radially shearing mesh.  Computational zones can move and shear radially, so that each radial track behaves much like a one-dimensional Lagrangian code, and the different tracks are coupled laterally by transverse fluxes.  Some additional details on the technique are presented in \cite{2013ApJ...775...87D}.  Mesh motion can substantially reduce errors from advecting the turbulence across the computational domain.  Another advantage of the moving-mesh technique is that the inner and outer boundaries follow the blast, so that the flow can move over a large dynamic range, while the code only needs to cover a modest dynamic range at any single time.  Multidimensional calculations have a typical resolution of $\Delta \theta \sim \Delta r / r \approx 4 \times 10^{-4}$.  In the appendix, a resolution study is performed demonstrating that this resolution is more than sufficient to capture the coarse-grained properties of the turbulence.  Mesh refinement and de-refinement is employed so that computational zones are typically kept at an aspect ratio close to unity.  1D calculations are performed using a custom-built 1D moving-mesh hydro code.  The 1D source code is publicly available at \texttt{https://github.com/duffell/RT1D}.

The outer boundary condition is Dirichlet, setting the values at this boundary equal to the initial condition.  This is appropriate as the outer boundary always consists of unshocked material, which is stationary until the ejecta collides with it.  Mesh refinement and de-refinement is performed based on the aspect ratio of zones.  If a zone has an aspect ratio higher than 5:1 or lower than 1:5, the zone is either split in half, or merged with its neighbor.  In either case, this action is performed in such a way as to conserve mass, energy, and momentum during the refinement or de-refinement process.  In principle, it is possible to seed the instability with an initial perturbation.  In practice, it has been found that the nonlinear phase is insensitive to the initial seed field, so long as this perturbation is small \citep[see also][]{2013ApJ...775...87D}.  This study works in the limiting case where the initial perturbation is arbitrarily small, and the instability is seeded by grid noise.  The fact that the coarse-grained end state is converged (see Appendix) demonstrates that the final state is indeed insensitive to the magnitude of the initial perturbation.

The first test, the ``Idealized Supernova" is used to calibrate the 1D model, determining appropriate choices of constants $A$, $B$, and $C$ by comparing with a very well-resolved 2D calculation.  This test will also be used to measure the effectiveness of the model, but these results should be taken with a grain of salt since the model was calibrated to perform as well as possible on this problem.

The ``Soft Explosion" is a test which is designed to produce smoother, less dramatic shocks, to facilitate comparison with numerical schemes which are not as robust at shock capturing.  This initial condition is smooth everywhere and the pressure is non-negligible so that the shocks generated are not as strong.  The contact wave in this case is not a sharp discontinuity, and can be easily resolved.

The ``Sustained Engine" starts with a static initial condition, and deposits thermal energy on a non-negligible timescale.  This test provides several independent checks on the model, including the fact that the engine generates an RT-unstable region from the opposite direction (negative pressure gradient and positive density gradient).  This test also includes a parameter which controls the steepness of the contact discontinuity, which will be varied to further test the robustness of the model.

The ``Brief Encounter" is a somewhat more complicated initial condition, used to test the effectiveness of the model in less idealized scenarios.  The ejecta density in this model obeys a broken power-law in radius, instead of a purely uniform ejecta model, and the CSM is described by a shell of material that is only encountered by the ejecta for a brief amount of time.  This test and the final test are also calculated in cgs units, although it should be stressed that the source terms and diffusive fluxes were introduced in a scale-invariant way, and therefore the method would perform identically in non-dimensional ``code units". 

The final test, the W7 model, uses a very nontrivial initial condition for the ejecta that is read from a table.  This test is meant to demonstrate performance on a pragmatic, complex problem that is not built out of simple analytic functions.

\subsection{Test \#1: The Idealized Supernova}
\label{sec:test1}

\begin{figure}
\epsscale{1.2}
\plotone{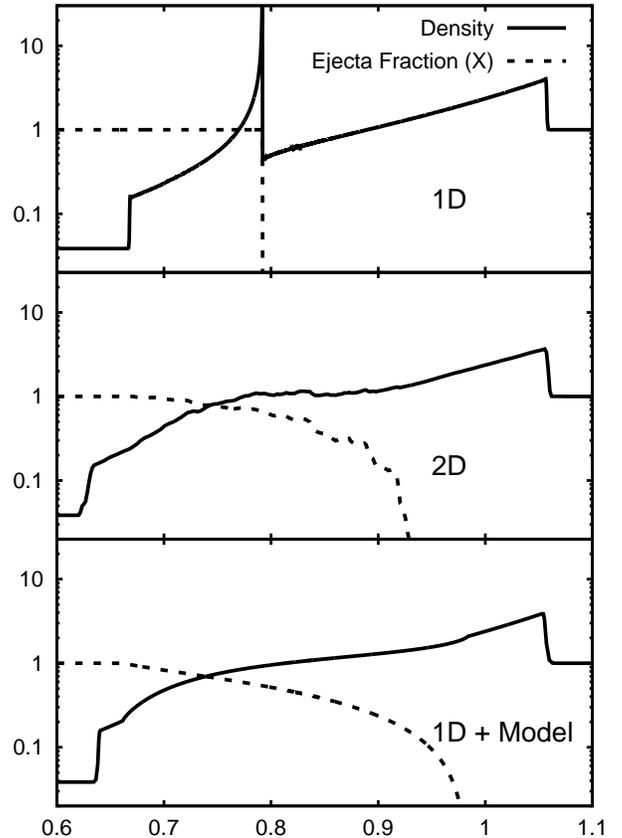}
\caption{ 1D angle-averaged profiles of the idealized supernova.  In this figure, density is plotted alongside the passive scalar X.  This passive scalar is designed to measure mixing of ejecta with CSM ($X=1$ for ejecta, and $X=0$ in the CSM).
\label{fig:idealsn2} }
\end{figure}

\begin{figure}
\epsscale{1.2}
\plotone{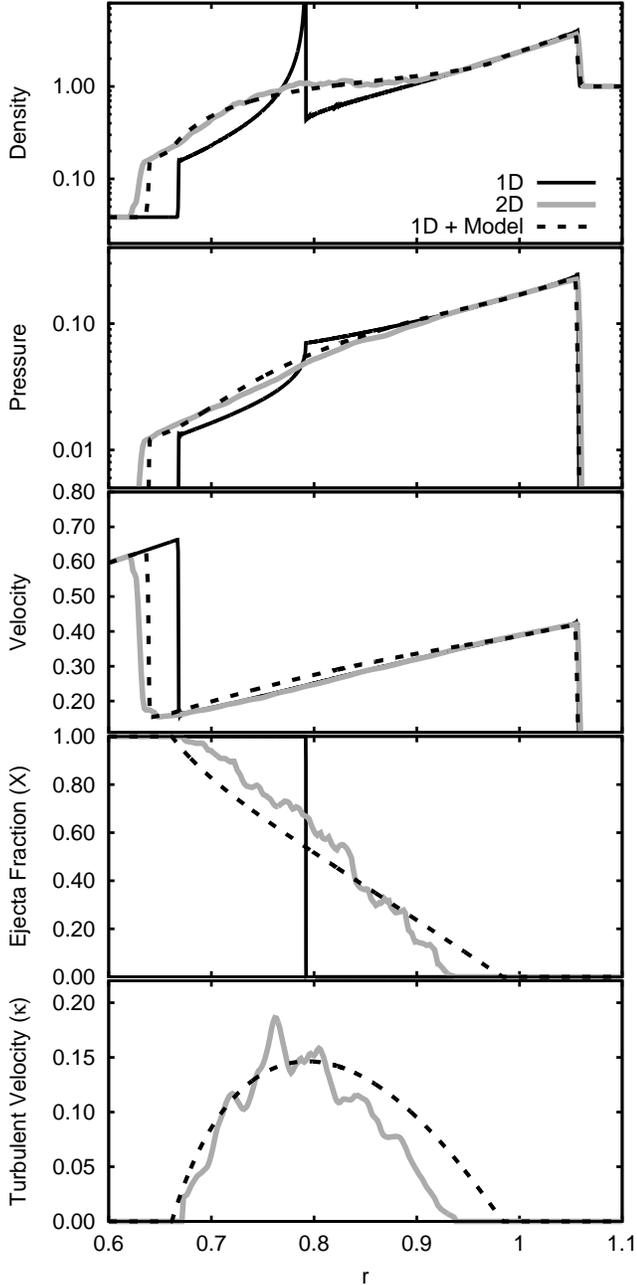}
\caption{ 1D angle-averaged profiles of the idealized supernova at time $t = 1.0$.  All variables are plotted for ease of comparison between 1D, 2D and the 1D model introduced in this work.  The most dramatic multidimensional effects are the seen in the density and passive scalar.
\label{fig:idealsn3} }
\end{figure}

\begin{figure}
\epsscale{1.2}
\plotone{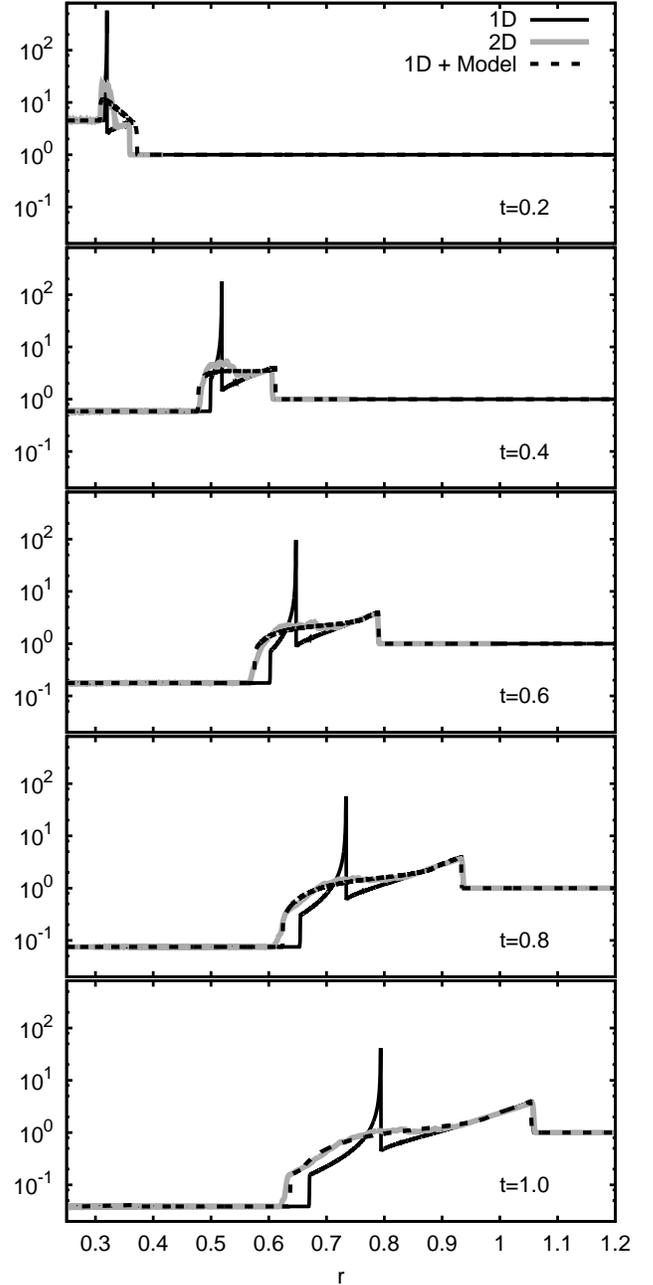}
\caption{ Snapshots of the idealized supernova at various times.  At early times ($t = 0.2$), the coarse-grained dynamics are only crudely approximated by the 1D model.  This is most likely due to the fact that equation (\ref{eqn:grow2}) is used for the growth rates, whereas the linear growth rate would probably be more appropriate at early times.  Nevertheless, the position of the reverse shock is accurately reproduced at all times, and the disruption of the extremely sharp contact discontinuity is reasonably well approximated.
\label{fig:times} }
\end{figure}

\begin{figure}
\epsscale{1.2}
\plotone{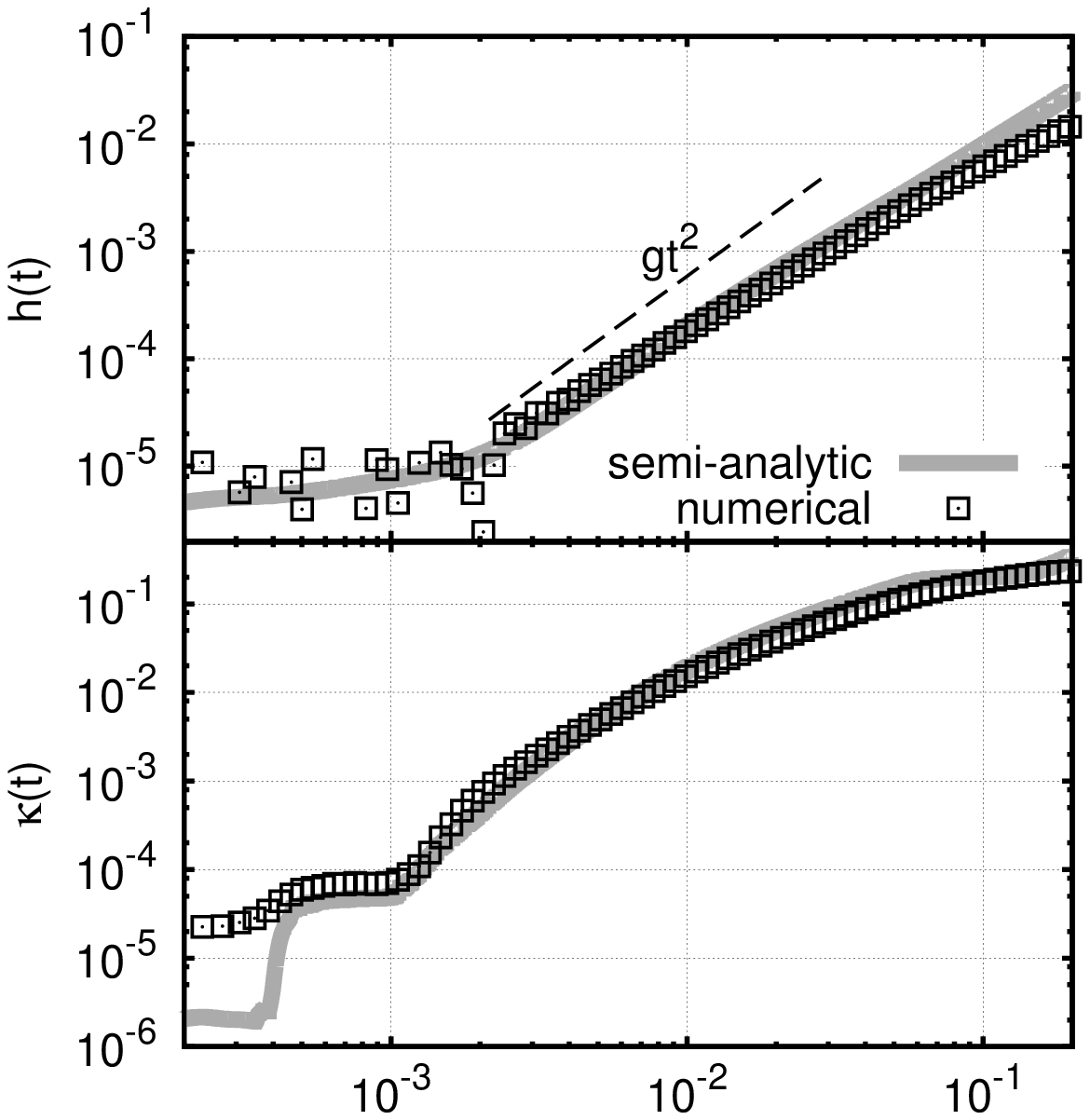}
\caption{ Growth of the instability at early times is compared with analytical predictions.  Given the acceleration $g$ as a function of time, semi-analytical curves are drawn for $h(t)$ and $\kappa(t)$ using equations (\ref{eqn:hanly}) and (\ref{eqn:kanly}), respectively.  These are compared with the numerically measured $h(t) = \sqrt{\left<r^2\right> - \left<r\right>^2}$ and $\left<\kappa(t)\right>$, where brackets denote an average over the mixing region (equation \ref{eqn:wdef}).
\label{fig:h} }
\end{figure}

This is the simplest RT-unstable supernova model that one can write down.  A uniform-density ball of gas expands freely into a uniform-density medium.  Initial conditions are given by 

\begin{equation}
\rho = \left\{ \begin{array}
				{l@{\quad \quad}l}
				1/(4 \pi R^3 / 3) & r < R \\
				1 & r > R
    			\end{array} \right.
\end{equation}

\begin{equation}
v = \left\{ \begin{array}
				{l@{\quad \quad}l}
				v_{max} (r/R) & r < R \\
				0 & r > R
    			\end{array} \right.
\end{equation}

and the pressure is negligible:

\begin{equation}
P = 10^{-5} \rho v_{max}^2.
\end{equation}

The remaining constants are $R = 0.01$, $v_{max} = \sqrt{10/3}$.  These initial conditions are chosen to give the ejecta a total mass $M = 1$ and energy $E = 1$, and to have it collide with a uniform CSM of density $\rho_{\rm CSM} = 1$.  The solution is then evaluated at time $t=1$, when the ejecta and CSM masses are comparable, and the RT instability is prominent (Figure \ref{fig:idealsn1}).

Figure \ref{fig:idealsn2} shows 1D angle-averaged values of $\rho$ and $X$ as a function of radius.  The passive scalar is plotted to show how much mixing is present between ejecta and CSM.  In the 1D case, there is of course no mixing.  The 2D case, however, shows a great deal of mixing between ejecta and CSM, completely smoothing away the sharp density jump seen in 1D.  The 1D RT model captures this mixing extremely well, though again this is not too surprising, as the constants $A$, $B$, and $C$ were found by trying to match this particular solution as closely as possible.  At $r = 0.85$, the blastwave is mixed with $36\%$ ejecta according to the 2D results, and $37\%$ ejecta according to the 1D model.  The model predicts slightly more penetration of the RT fingers (the passive scalar $X$ is nonzero over a larger range in the model), but is otherwise in agreement with the 2D results.

For a more complete comparison, Figure \ref{fig:idealsn3} plots all quantities as a function of radius for the 1D, 2D and ``1D+Model" calculations.  The most dramatic differences between 1D and 2D are seen in the density $\rho$ and the ejecta fraction $X$, which is why those two quantities have been chosen to represent the results in most figures.  The turbulent velocity represented by the quantity $\kappa$ is also plotted in the bottom panel of Figure \ref{fig:idealsn3}.  This quantity is compared directly with the 2D results using equation (\ref{eqn:kappa0}) to calculate $\kappa$, and evaluating the kinetic turbulent energy density in the same manner as in \cite{2013ApJ...775...87D}, but in the nonrelativistic limit.  Putting these results together, $\kappa$ is computed via the formula:

\begin{equation}
\kappa = {2 \over \gamma (\gamma-1)} { \langle P \rangle_{\rm cons} - \langle P \rangle_{\rm vol} \over \langle P \rangle_{\rm vol} }
\end{equation}

where brackets denote an average over a spherical shell, $\langle \rangle_{vol}$ denotes a volume average, while $\langle \rangle_{cons}$ denotes a ``conservative average", where the total mass, energy, and radial momentum of the shell are calculated and these are converted back to the primitive variables of density, pressure, and velocity.  A full derivation of the relativistic version of this formula is presented in \cite{2013ApJ...775...87D}.

The constants $A$, $B$, and $C$ have been tuned so that the profile of $\kappa$ matches the 2D results as closely as possible.  These choices of $A = 1.18 \times 10^{-5}$, $B = 1.2$, and $C = 0.102$ are fixed when applied to the remaining test problems (sections \ref{sec:testx} - \ref{sec:test3}).

Due to the inherent scale-invariance of hydrodynamics, a great deal of parameter space is already implicitly explored in this single test.  Performing the same test with a different value of $E$, $M$, or $\rho_{\rm CSM}$ would yield identical results, assuming the limit $R \ll (M/\rho_{\rm CSM})^{1/3}$.  Since the 1D model is introduced in a scale-invariant way, this model would also perform identically well for an idealized supernova with any other choice of $E$, $M$, or $\rho_{\rm CSM}$.  This is assuming the snapshot is taken at the re-scaled time

\begin{equation}
t = t_0 = M^{5/6} E^{-1/2} \rho_{\rm CSM}^{-1/3}.
\end{equation}

Therefore, the only true free parameter in this test problem is the time at which the solution is evaluated (i.e. the dimensionless parameter $t/t_0$).  For additional free parameters, one could also vary the detailed profile of the ejecta or CSM, but that is precisely what will be tested in the remaining four test problems (sections \ref{sec:testx} - \ref{sec:test3}).

Figure \ref{fig:times} presents the idealized supernova taken at several different snapshots in time, to explore how well the model captures the evolution of the instability, from early times when the density jump is very large, into late times when all of the gradients are smoothed out and shallow.  At very early times ($t = 0.2$), the 1D model only crudely approximates the solution, capturing the disruption of the extremely narrow contact discontinuity in an order-of-magnitude sense.  Inaccuracies here may have to do with the fact that equation (\ref{eqn:grow2}) is used for the growth rate, which is possibly not valid at early times.

Nevertheless, after this brief transient phase, the 1D model rapidly approaches the correct 2D solution, capturing the position of the reverse shock accurately, as well as the disruption of the contact discontinuity.

\hilight{Finally, the initial growth phase of the instability is tested against analytical predictions for the evolution of width of the mixing layer $h(t)$, and for $\kappa(t)$.  Equation (\ref{eqn:hnonl}) for the growth of the instability suggests that the width of the mixing layer obeys}

\begin{equation}
\ddot h = 2 \alpha g \mathcal{A},
\end{equation}

\hilight{where $g = P'/\rho$ is the acceleration felt by the mixing layer.  Therefore, $h(t)$ can be predicted to follow}

\begin{equation}
h(t) = \alpha \int_0^t dt' \int_0^{t'} dt'' 2 \mathcal{A} g(t'').
\label{eqn:hanly}
\end{equation}

\hilight{For $\kappa$, one can assume the growth rate of the mixing layer is proportional to the turbulent velocity fluctuation:}

\begin{equation}
\dot h \propto \delta u,
\end{equation}

\hilight{for some constant of proportionality.  In this case, one can readily predict that}

\begin{equation}
\kappa \propto \left( \int_0^t dt' \mathcal{A} g(t') \right)/c^2.
\label{eqn:kanly}
\end{equation}

\hilight{The quantities $\left<g\right>$, $\left<\kappa\right>$, $\left<c\right>$, and the width of the mixing layer $h$ are measured as averages in the mixing region, using the weighting function}

\begin{equation}
w(r) \equiv X(r) ( 1 - X(r) ),
\end{equation}

\hilight{which only has support in the mixed region.  Averaged quantities are calculated via}

\begin{equation}
\left<Q\right>_w = { \int w(r) Q(r) dr \over \int w(r) dr },
\label{eqn:wdef}
\end{equation}

\hilight{and the mixing layer thickness is given by the standard deviation of $w$:}

\begin{equation}
h = \sqrt{ \left<r^2\right>_w - \left<r\right>_w^2 }.
\end{equation}

\hilight{$h(t)$ and the averaged $\left<\kappa(t)\right>_w$ are plotted as a function of time in Figure \ref{fig:h}.  These are plotted against computed values integrating equations (\ref{eqn:hanly}) and (\ref{eqn:kanly}), given the acceleration $\left<g(t)\right>_w$ as a function of time (for simplicity, it is assumed that $\mathcal{A} = 1$ as the density jump is very large at early times).  Growth produced by the model is consistent with nonlinear RT, with $\alpha = 0.2$.  This $\alpha$ is significantly larger than typical values, which is further evidence that the early-time growth is too fast (though the definition of $h$ could come with an overall dimensionless coefficient).  $\kappa(t)$ also appears consistent with (\ref{eqn:kanly}).}

\subsection{Test \#2: The Soft Explosion}
\label{sec:testx}

\begin{figure}
\epsscale{1.2}
\plotone{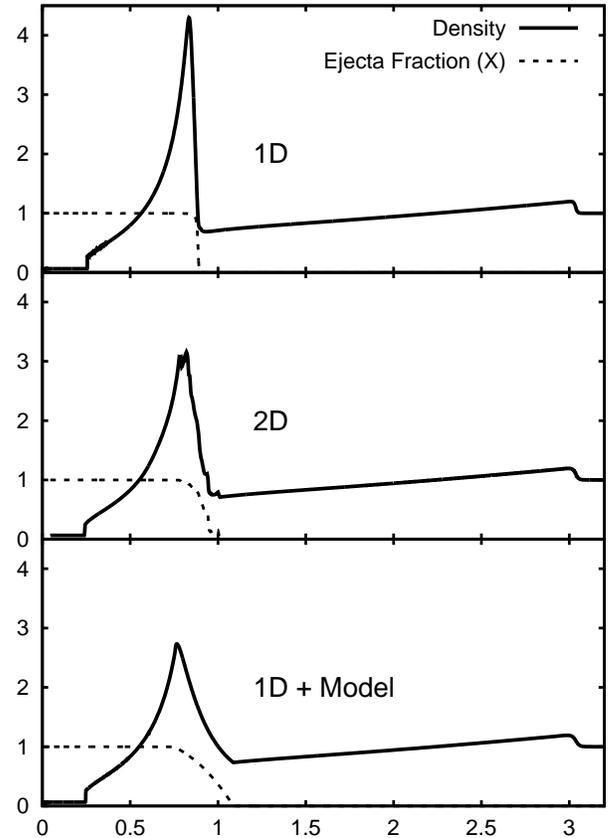}
\caption{ 1D angle-averaged profiles of the soft explosion at $t = 1.0$.  Shocks are weaker and the solution is smoother than the previous test.  RT mixing is also not as significant for this test problem, and this fact is faithfully recovered in the model.  The density peak is reduced from $\rho_{\rm max} = 4.3$ to $\rho_{\rm max} \approx 3$.
\label{fig:soft} }
\end{figure}

This is another simple set-up, but with smooth initial conditions that generate weaker shocks.  This is designed for comparison with other numerical implementations, for which shock capturing may be more challenging.  The shocks and density jump have very different strengths from the first test, so that this probes a significantly distinct region of parameter space.  Initial conditions are given by the following:

\begin{equation}
\rho(r) = \rho_{\rm ej}(r) + \rho_{\rm CSM}(r)
\end{equation}
\begin{equation}
P(r) = 2.5 \rho(r)
\end{equation}
\begin{equation}
X(r) = \rho_{\rm ej}(r) / \rho(r)
\end{equation}
\begin{equation}
v(r) = (r/R_0) X(r),
\end{equation}

where $R_0 = 0.1$ and

\begin{equation}
\rho_{\rm ej}(r) = e^{-(r/R_0)^2}/R_0^3,
\end{equation}
\begin{equation}
\rho_{\rm CSM}(r) = 1.
\end{equation}

The initial conditions are thus completely specified by smooth functions of radius, and the non-negligible pressure will cause the shocks to be weaker and more easily resolved.

Solutions for these initial conditions are presented in Figure \ref{fig:soft}.  For this test problem, the influence of RT is not as significant as in the previous test, and the model correctly captures this, and does not overmix.  The density peak in 1D is at about $\rho_{\rm max} = 4.3$, whereas RT instability reduces the density peak to $\rho_{\rm max} \approx 3$.  In this case, the reverse shock is relatively unaffected by the turbulence.

\subsection{Test \#3: The Persistent Engine}
\label{sec:testx2}

This test is significantly distinct from the others, as the energy is input to the system on a non-negligible timescale, modeling an engine which is persistent.  This has the effect of producing a much more complicated ejecta structure, and therefore this is a much ``messier" test than the previous ones, but more importantly the interior pressure causes an RT instability which is ``upside-down" with respect to the others; there is a negative pressure gradient coinciding with a positive density gradient.  In other words, low density material is \emph{accelerating} the high-density material in front of it, in contrast with the low-density material \emph{decelerating} high-density material behind it, as is the case in all of the other tests.  In addition, there will be the usual RT caused by the deceleration of the ejecta, so that the ejecta is unstable from both sides.  This test will help to demonstrate whether this model is reasonable for capturing different scenarios which can cause RT, other than a decelerating shell of ejecta.

Initial conditions are given by a static fluid of constant pressure:

\begin{equation}
P = 0.1
\end{equation}
\begin{equation}
v = 0
\end{equation}
\begin{equation}
\rho = 1 + { \rm{tanh}((R-r)/a) \over 2 R^3}
\end{equation}

where $R=0.1$ and $a$ is a free parameter which controls the steepness of the contact discontinuity and therefore the growth of RT.

Energy is deposited into the system thermally, as a source term.  The power per unit volume is given by:

\begin{equation}
S_{\epsilon} = {1 \over 4 \pi R_E^3 T /3 },~ r<R_E ~\rm{and} ~t<T
\end{equation}

where $R_E = 0.05$ and $T = 0.1$.  This results in an outflow with total energy $E = 1$ when the engine has terminated.

\begin{figure}
\epsscale{1.2}
\plotone{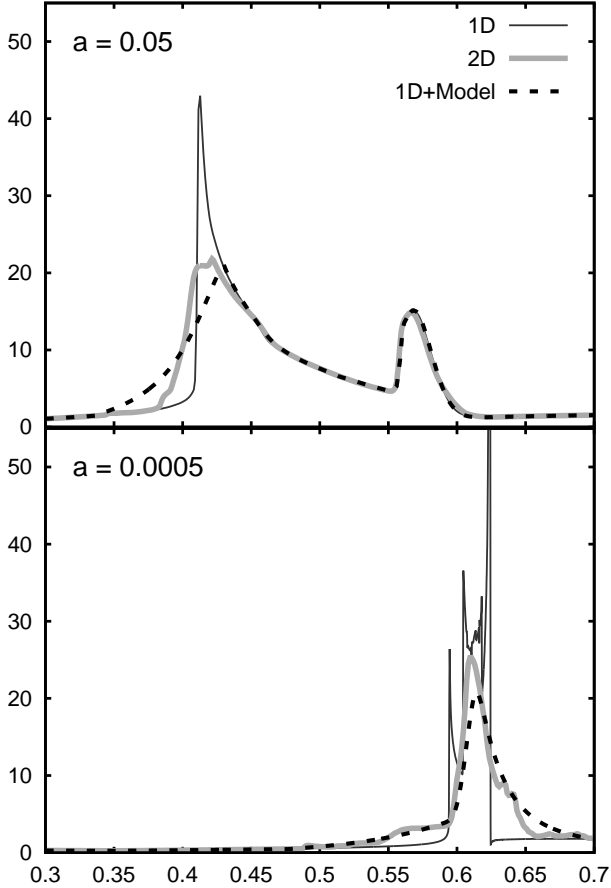}
\caption{ 1D angle-averaged profiles of the persistent engine test at $t = 1.0$.  Solutions are given for two different values of the parameter $a$, which measures the thickness of the transition between ejecta and CSM, setting the steepness of the density gradient in the unstable region.
\label{fig:persist} }
\end{figure}

Solutions are presented in Figure \ref{fig:persist} for two different values of the parameter $a$.  For broad initial conditions with $a = 0.05$, the density gradients are shallow and RT is weak; the internal RT caused by the engine reduces the density peak by about a factor of two, and the external RT caused by deceleration has very little effect on the solution.  This is reproduced in the 1D model.

The more extreme, ``sharp" jump with $a = 0.0005$ (lower panel of Figure \ref{fig:persist}) results in complex, fine structure in the 1D solution (note: these sharp ups and downs are not numerical; they are resolved and fully converged in 1D).

For this value of $a$, RT is much stronger and diffuses the entire region.  RT is prominent from both the interior and exterior of the ejecta, and it smoothes out the solution so that the density peaks at around $\rho \sim 20$ instead of sharply peaking at $\rho \sim 50$ as in the 1D version.  It is reassuring that all of these features are faithfully captured in the model, as the steepness of the contact discontinuity is varied by several orders of magnitude.

\subsection{Test \#4: The Brief Encounter}
\label{sec:test2}

\begin{figure}
\epsscale{1.2}
\plotone{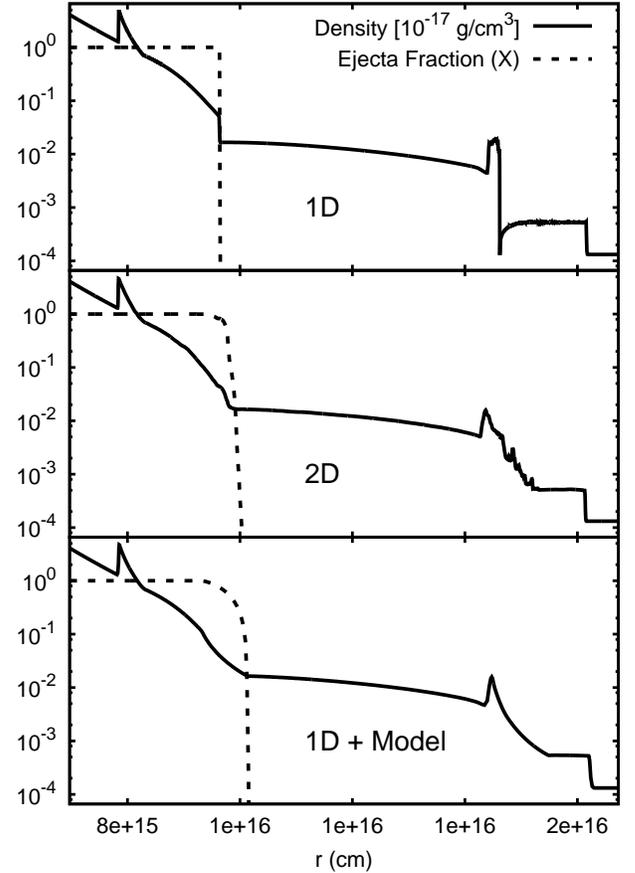}
\caption{ 1D angle-averaged profiles of the brief encounter at $t = 50$ days.  Similar to  Figures \ref{fig:idealsn2} and \ref{fig:soft}, density is plotted alongside the passive scalar X.  The multidimensional solution finds two distinct regions that are RT-unstable, associated with the two contact discontinuities in the initial condition.  The RT model presented here is able to capture both RT-unstable regions.
\label{fig:encounter} }
\end{figure}

This is a slightly more complicated initial condition.  The ejecta is given by a broken power-law, such that most of the ejected mass is contained within the radius $R_1$, but RT can be seen at early times, before the reverse shock passes this radius.  Additionally, the encounter does not last forever, as this is meant to represent a finite shell of gas that the supernova collides with.  The initial conditions are

\begin{equation}
\rho = \left\{ \begin{array}
				{l@{\quad \quad}l}
				\rho_c (R_1/R_c)^{-10} (r/R_1)^{-1} & r < R_1 \\
				\rho_c (r/R_c)^{-10} & R_1 < r < R_c \\
				\rho_{\text{CSM}} & R_c < r < R_2 \\
				10^{-4} \rho_{\text{CSM}} & r > R_2
    			\end{array} \right.
\end{equation}

\begin{equation}
v = \left\{ \begin{array}
				{l@{\quad \quad}l}
				r/t_0 & r < R_c \\
				0 & r > R_c
    			\end{array} \right.
\end{equation}

where $R_1 = 4.4 \times 10^{14}$ cm, $R_c = 1.4 \times 10^{15}$ cm, $R_2 = 2.8 \times 10^{15}$ cm, $\rho_{\text{CSM}} = 1.3 \times 10^{-17}$ g/cm$^3$, $\rho_c = 4 \times 10^{-17}$ g/cm$^3$, $t_0 = 5$ days.  For this problem, the temperature is negligible:

\begin{equation}
T = 10^4 K
\end{equation}

The solution is evaluated at $t = 45$ days later, a total of $t + t_0 = 50$ days after the supernova (Figure \ref{fig:encounter}).  At this time, the shocks have only encountered the steep $r^{-10}$ density gradient, but the forward shock has overtaken the entirety of the shell.  After crossing the shell, a second forward shock is pushed forward into the lower-density ``vacuum" state.  A 2D study of this solution shows that two separate regions are RT-unstable, associated with the two different contact discontinuities in the initial condition.  The first unstable region corresponds to the inner edge of the shell, being disrupted by RT as the shell decelerates the ejecta.  The second unstable region applies to the interface between the outer edge of the shell and the ``vacuum" state just outside; this density drop is also unstable.  These two separate unstable regions are faithfully reproduced in the 1D model presented in this work, again using the same parameters calibrated by the first test.

\subsection{Test \#5: The W7 Model}
\label{sec:test3}

\begin{figure}
\epsscale{1.2}
\plotone{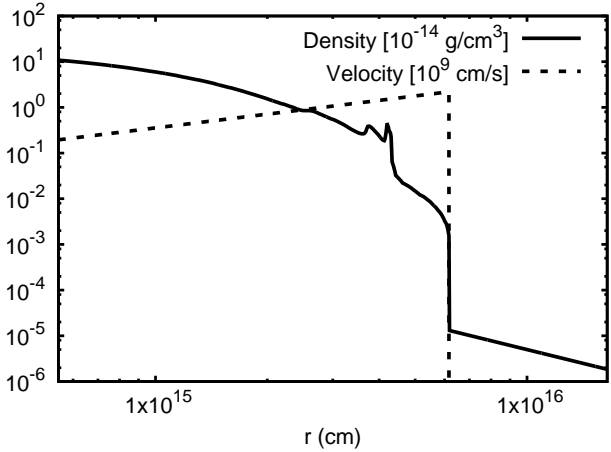}
\caption{ Initial conditions for the W7 Model.  This final test was read from a table, rather than being built out of simple analytic functions, in contrast with the previous tests.
\label{fig:w7init} }
\end{figure}

\begin{figure}
\epsscale{1.2}
\plotone{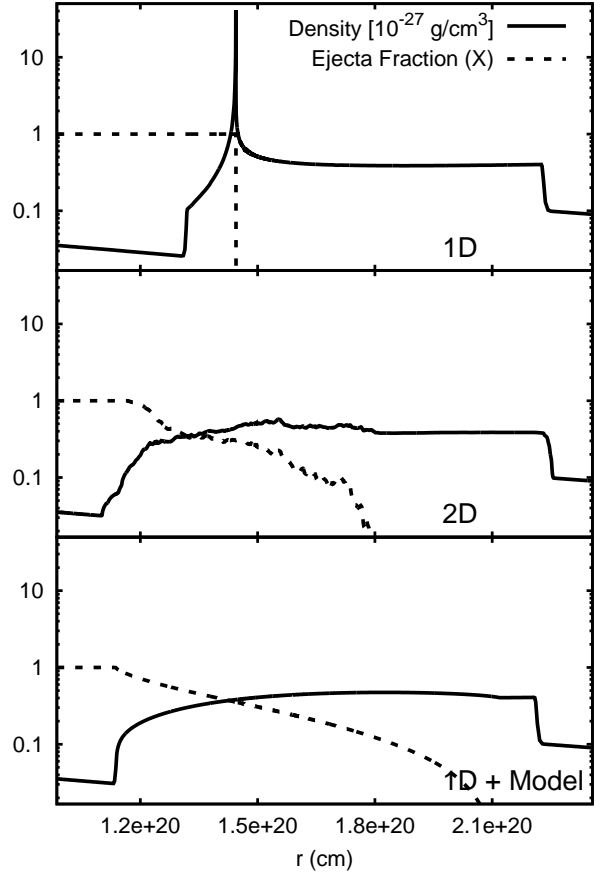}
\caption{ W7 Model at time $t = 1187$ years.  At this late time, the supernova remnant has swept up a signficant portion of the CSM.
\label{fig:w7} }
\end{figure}

W7 is a commonly-used pure deflagration type-1a supernova ejecta model that was calculated using detailed nucleosynthesis \citep{1984ApJ...286..644N}.  The data is publicly available and therefore makes an ideal test case for complicated supernova ejecta profiles.  In this example, the ejecta is assumed to be expanding into a wind, with density

\begin{equation}
\rho_{wind}(r) = K/r^2
\end{equation}

with $K = 5 \times 10^{12}$ g/cm.  This wind corresponds to a mass loss rate of $10^{-4} M_{\sun}$ per year and a wind velocity of $1000$ km/s.  The initial density and velocity profile is plotted in Figure \ref{fig:w7init}.

Results from the W7 model are shown in Figure \ref{fig:w7}.  The supernova remnant is shown at very late times, $t = 1187$ years, after which a substantial amount of the flow has been decelerated by the wind (it might be more realistic to use a shallower density profile at such large radii, but this is an unimportant detail for the purposes of the current study).  The 1D solution is characterized by a sharp density jump that is unstable in 2D.  Again, the model presented in this work accurately captures the 2D effects, and provides a much more accurate approximation to the 2D solution than the 1D version.

In the 1D case, there is again a sharp contact discontinuity, but in 2D, there is significant mixing between ejecta and CSM.  This also pushes the reverse shock significantly backward.  All of these basic features are reproduced by the 1D model, again using the same parameters $A$, $B$, and $C$ which were calibrated by the first test.  The main distinction is again the penetration depth of the RT fingers.  At radii up to $r \sim 1.7 \times 10^{20}$ cm, the model predicts the correct amount of mixing (about $10\%$ of ejecta mixed with CSM), but at larger radii the model somewhat overpredicts how deeply the fingers penetrate when compared to the 2D results.

\section{Summary}
\label{sec:summary}

\begin{figure}
\epsscale{1.2}
\plotone{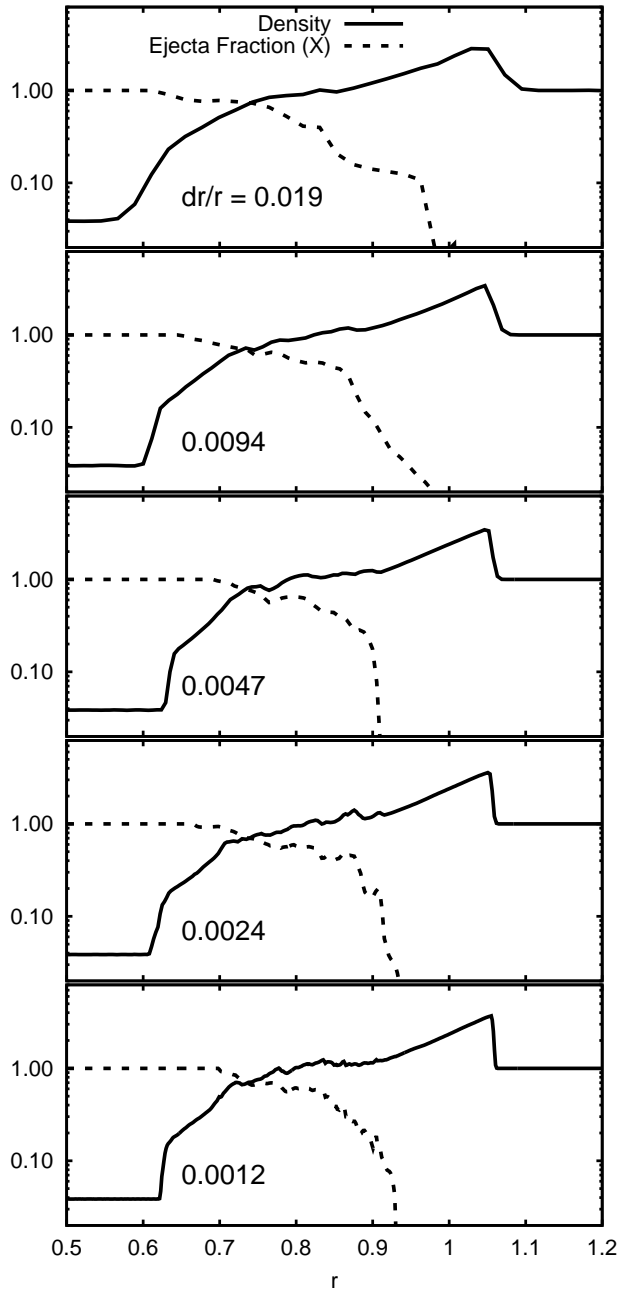}
\caption{ Test \#1 is performed at five different resolutions in 2D, in order to test convergence.  In all cases, $\delta \theta \sim \delta r / r$.
\label{fig:res_study} }
\end{figure}

\hilight{Connecting supernova models to observations may require extensive searches of parameter space, which may only be possible in 1D, especially if these supernova models include full radiation hydrodynamics, which is expensive in multidimensions.  The model presented here provides a useful tool for exploring this parameter space.  This is potentially a big improvement over the current standard, which is to perform the calculation in 1D, then manually smooth out offending regions in post-processing \citep[][]{1988ApJ...329..820P, 2009ApJ...703.2205K, 2010ApJ...724..341H}.}

Since the modified equations give a value for $\kappa = \delta u^2 / c_s^2$, this potentially provides an estimate for the magnetic field strength, if magnetic fields are assumed to be in equipartition with kinetic fluctuations \citep{2003ApJ...597L.141H, 2004ApJ...612..276S,  2012PhRvL.108c5002B, 2013ApJ...769L..29Z, 2013ApJ...775...87D, 2014ApJ...791L...1D}.  In this case, $\epsilon_B \sim (\gamma(\gamma-1)/2) \kappa$, by Equation (\ref{eqn:kappa0}).

The current model is accurate enough to be used immediately in 1D studies of supernovae, but it should be stressed that this is only a first step (actually a second step, after \cite{1973MNRAS.161...47G}) toward a complete RT model.  It should also be noted that the model includes several fitting parameters, and that this might not be a unique method of matching the multidimensional results.

\hilight{The model is designed and calibrated to match late-time behavior, after initial transients have washed out and the swept-up mass is comparable to the mass of the ejecta.  However, at very late times, the decay of turbulence is important, and this has not yet been accounted for.  A more accurate model would utilize a linear model for the early-time growth rate, and include the decay term to match late-time evolution.  A connection between the early and late phases of growth has been realized in the 1D ODE models of \citep{ramshaw1998simple}, and this model was incorporated in a multi-phase PDE mixing model \citep{ramshaw2000implementation}, so it is likely possible to incorporate such modifications into the single-fluid PDE model presented here in a future study.}

Another important point is that the constants $A,B,C$ were calibrated using 2D numerical results.  3D results would be much more accurate, even if they are not resolved as well as their 2D counterparts.  It would also be desirable to extend this model to relativistic flows, for application to gamma ray burst afterglows.  Finally, it is well-known that cooling from cosmic rays can strongly impact the dynamics of RT, and it is unclear whether this would need to be accounted for in the model.  This should be tested explicitly in a future study.

\acknowledgments 

Resources supporting this work were provided by the NASA High-End Computing (HEC) Program through the NASA Advanced Supercomputing (NAS) Division at Ames Research Center.  I am grateful to Chris McKee, Dan Kasen, Eliot Quataert, Andrei Gruzinov, and Nathan Roth for helpful comments and discussions.  I would especially like to thank Chelsea Harris for encouraging me to attempt this study in the first place, and Bill Paxton for additional encouragement and for inspiring several of the tests performed in this study.  I would also like to thank the anonymous referee for the extremely detailed and thorough review.

\begin{appendix}
\section{Resolution Study}

In order to demonstrate that the tests in this study are based on converged numerical solutions, a resolution study is performed on the first test (the ``Idealized Supernova", see section \ref{sec:test1} and Figure \ref{fig:idealsn1}).

This test is performed in 2D at five different resolutions, shown in Figure \ref{fig:res_study}.  Even the lowest resolution $\Delta r / r \sim 0.02$ captures the coarse-grained effects reasonably well (surprisingly so).  As the resolution is increased,  the angle-averaged quantities measured in this study approach a converged solution.

\end{appendix}

\bibliographystyle{apj}


\begin{thebibliography}{}
\expandafter\ifx\csname natexlab\endcsname\relax\def\natexlab#1{#1}\fi

\bibitem[{{Alon} {et~al.}(1994){Alon}, {Hecht}, {Mukamel}, \&
  {Shvarts}}]{1994PhRvL..72.2867A}
{Alon}, U., {Hecht}, J., {Mukamel}, D., \& {Shvarts}, D. 1994, Physical Review
  Letters, 72, 2867

\bibitem[{Ashurst \& Kerstein(2005)}]{ashurst2005one}
Ashurst, W.~T., \& Kerstein, A.~R. 2005, Physics of Fluids (1994-present), 17,
  025107

\bibitem[{{Beresnyak}(2012)}]{2012PhRvL.108c5002B}
{Beresnyak}, A. 2012, Physical Review Letters, 108, 035002

\bibitem[{{Blondin} \& {Ellison}(2001)}]{2001ApJ...560..244B}
{Blondin}, J.~M., \& {Ellison}, D.~C. 2001, \apj, 560, 244

\bibitem[{{Budil} {et~al.}(1996){Budil}, {Remington}, {Peyser}, {Mikaelian},
  {Miller}, {Woolsey}, {Wood-Vasey}, \& {Rubenchik}}]{1996PhRvL..76.4536B}
{Budil}, K.~S., {Remington}, B.~A., {Peyser}, T.~A., {et~al.} 1996, Physical
  Review Letters, 76, 4536

\bibitem[{{Casner} {et~al.}(2012){Casner}, {Smalyuk}, {Masse}, {Igumenshchev},
  {Liberatore}, {Jacquet}, {Chicanne}, {Loiseau}, {Poujade}, {Bradley}, {Park},
  \& {Remington}}]{2012PhPl...19h2708C}
{Casner}, A., {Smalyuk}, V.~A., {Masse}, L., {et~al.} 2012, Physics of Plasmas,
  19, 082708

\bibitem[{{Chandrasekhar}(1961)}]{1961hhs..book.....C}
{Chandrasekhar}, S. 1961, {Hydrodynamic and hydromagnetic stability}

\bibitem[{{Chevalier}(1982)}]{1982ApJ...258..790C}
{Chevalier}, R.~A. 1982, \apj, 258, 790

\bibitem[{{Chevalier} {et~al.}(1992){Chevalier}, {Blondin}, \&
  {Emmering}}]{1992ApJ...392..118C}
{Chevalier}, R.~A., {Blondin}, J.~M., \& {Emmering}, R.~T. 1992, \apj, 392, 118

\bibitem[{{Chevalier} \& {Klein}(1978)}]{1978ApJ...219..994C}
{Chevalier}, R.~A., \& {Klein}, R.~I. 1978, \apj, 219, 994

\bibitem[{{Dimonte} \& {Schneider}(2000)}]{2000PhFl...12..304D}
{Dimonte}, G., \& {Schneider}, M. 2000, Physics of Fluids, 12, 304

\bibitem[{{Dimonte} {et~al.}(2004){Dimonte}, {Youngs}, {Dimits}, {Weber},
  {Marinak}, {Wunsch}, {Garasi}, {Robinson}, {Andrews}, {Ramaprabhu}, {Calder},
  {Fryxell}, {Biello}, {Dursi}, {MacNeice}, {Olson}, {Ricker}, {Rosner},
  {Timmes}, {Tufo}, {Young}, \& {Zingale}}]{2004PhFl...16.1668D}
{Dimonte}, G., {Youngs}, D.~L., {Dimits}, A., {et~al.} 2004, Physics of Fluids,
  16, 1668

\bibitem[{Dimonte {et~al.}(2004)Dimonte, Youngs, Dimits, Weber, Marinak,
  Wunsch, Garasi, Robinson, Andrews, Ramaprabhu,
  {et~al.}}]{dimonte2004comparative}
Dimonte, G., Youngs, D., Dimits, A., {et~al.} 2004, Physics of Fluids
  (1994-present), 16, 1668

\bibitem[{{Duffell} \& {MacFadyen}(2011)}]{2011ApJS..197...15D}
{Duffell}, P.~C., \& {MacFadyen}, A.~I. 2011, \apjs, 197, 15

\bibitem[{{Duffell} \& {MacFadyen}(2013)}]{2013ApJ...775...87D}
---. 2013, \apj, 775, 87

\bibitem[{{Duffell} \& {MacFadyen}(2014)}]{2014ApJ...791L...1D}
---. 2014, \apjl, 791, L1

\bibitem[{{Ferrand} {et~al.}(2010){Ferrand}, {Decourchelle}, {Ballet},
  {Teyssier}, \& {Fraschetti}}]{2010AnA...509L..10F}
{Ferrand}, G., {Decourchelle}, A., {Ballet}, J., {Teyssier}, R., \&
  {Fraschetti}, F. 2010, \aap, 509, L10

\bibitem[{{Fraschetti} {et~al.}(2010){Fraschetti}, {Teyssier}, {Ballet}, \&
  {Decourchelle}}]{2010AnA...515A.104F}
{Fraschetti}, F., {Teyssier}, R., {Ballet}, J., \& {Decourchelle}, A. 2010,
  \aap, 515, A104

\bibitem[{{Glendinning} {et~al.}(2003){Glendinning}, {Bolstad}, {Braun},
  {Edwards}, {Hsing}, {Lasinski}, {Louis}, {Miles}, {Moreno}, {Peyser},
  {Remington}, {Robey}, {Turano}, {Verdon}, \& {Zhou}}]{2003PhPl...10.1931G}
{Glendinning}, S.~G., {Bolstad}, J., {Braun}, D.~G., {et~al.} 2003, Physics of
  Plasmas, 10, 1931

\bibitem[{Gonzalez-Juez {et~al.}(2013)Gonzalez-Juez, Kerstein, \&
  Lignell}]{gonzalez2013reactive}
Gonzalez-Juez, E., Kerstein, A., \& Lignell, D. 2013, Geophysical \&
  Astrophysical Fluid Dynamics, 107, 506

\bibitem[{{Gull}(1973)}]{1973MNRAS.161...47G}
{Gull}, S.~F. 1973, \mnras, 161, 47

\bibitem[{{Haan}(1989)}]{1989PhRvA..39.5812H}
{Haan}, S.~W. 1989, \pra, 39, 5812

\bibitem[{{Haugen} {et~al.}(2003){Haugen}, {Brandenburg}, \&
  {Dobler}}]{2003ApJ...597L.141H}
{Haugen}, N.~E.~L., {Brandenburg}, A., \& {Dobler}, W. 2003, \apjl, 597, L141

\bibitem[{{Heger} \& {Woosley}(2010)}]{2010ApJ...724..341H}
{Heger}, A., \& {Woosley}, S.~E. 2010, \apj, 724, 341

\bibitem[{Jozefik {et~al.}(2015)Jozefik, Kerstein, \&
  Schmidt}]{jozefik2015towards}
Jozefik, Z., Kerstein, A.~R., \& Schmidt, H. 2015, in Active Flow and
  Combustion Control 2014 (Springer), 197--211

\bibitem[{{Jun} \& {Norman}(1996)}]{1996ApJ...465..800J}
{Jun}, B.-I., \& {Norman}, M.~L. 1996, \apj, 465, 800

\bibitem[{{Jun} {et~al.}(1995){Jun}, {Norman}, \&
  {Stone}}]{1995ApJ...453..332J}
{Jun}, B.-I., {Norman}, M.~L., \& {Stone}, J.~M. 1995, \apj, 453, 332

\bibitem[{{Kane} {et~al.}(2000){Kane}, {Arnett}, {Remington}, {Glendinning},
  {Baz{\'a}n}, {M{\"u}ller}, {Fryxell}, \& {Teyssier}}]{2000ApJ...528..989K}
{Kane}, J., {Arnett}, D., {Remington}, B.~A., {et~al.} 2000, \apj, 528, 989

\bibitem[{{Kasen} \& {Woosley}(2009)}]{2009ApJ...703.2205K}
{Kasen}, D., \& {Woosley}, S.~E. 2009, \apj, 703, 2205

\bibitem[{{Kerstein}(1999)}]{1999JFM...392..277K}
{Kerstein}, A.~R. 1999, Journal of Fluid Mechanics, 392, 277

\bibitem[{{Kuranz} {et~al.}(2010){Kuranz}, {Drake}, {Grosskopf}, {Fryxell},
  {Budde}, {Hansen}, {Miles}, {Plewa}, {Hearn}, \&
  {Knauer}}]{2010PhPl...17e2709K}
{Kuranz}, C.~C., {Drake}, R.~P., {Grosskopf}, M.~J., {et~al.} 2010, Physics of
  Plasmas, 17, 052709

\bibitem[{{Lindl}(1995)}]{1995PhPl....2.3933L}
{Lindl}, J. 1995, Physics of Plasmas, 2, 3933

\bibitem[{{Lindl} \& {Mead}(1975)}]{1975PhRvL..34.1273L}
{Lindl}, J.~D., \& {Mead}, W.~C. 1975, Physical Review Letters, 34, 1273

\bibitem[{{Mac Low}(1999)}]{1999ApJ...524..169M}
{Mac Low}, M.-M. 1999, \apj, 524, 169

\bibitem[{{McKee}(1974)}]{1974ApJ...188..335M}
{McKee}, C.~F. 1974, \apj, 188, 335

\bibitem[{{Nomoto} {et~al.}(1984){Nomoto}, {Thielemann}, \&
  {Yokoi}}]{1984ApJ...286..644N}
{Nomoto}, K., {Thielemann}, F.-K., \& {Yokoi}, K. 1984, \apj, 286, 644

\bibitem[{{Ofer} {et~al.}(1996){Ofer}, {Alon}, {Shvarts}, {McCrory}, \&
  {Verdon}}]{1996PhPl....3.3073O}
{Ofer}, D., {Alon}, U., {Shvarts}, D., {McCrory}, R.~L., \& {Verdon}, C.~P.
  1996, Physics of Plasmas, 3, 3073

\bibitem[{{Oron} {et~al.}(2001){Oron}, {Arazi}, {Kartoon}, {Rikanati}, {Alon},
  \& {Shvarts}}]{2001PhPl....8.2883O}
{Oron}, D., {Arazi}, L., {Kartoon}, D., {et~al.} 2001, Physics of Plasmas, 8,
  2883

\bibitem[{{Pinto} \& {Woosley}(1988)}]{1988ApJ...329..820P}
{Pinto}, P.~A., \& {Woosley}, S.~E. 1988, \apj, 329, 820

\bibitem[{Ramshaw(2000)}]{ramshaw2000implementation}
Ramshaw, J. 2000, Tech. rep., Report UCRL-JC-139800, Lawrence Livermore
  National Laboratory

\bibitem[{Ramshaw(1998)}]{ramshaw1998simple}
Ramshaw, J.~D. 1998, Physical Review E, 58, 5834

\bibitem[{{Remington} {et~al.}(1999){Remington}, {Arnett}, {Drake}, \&
  {Takabe}}]{1999Sci...284.1488R}
{Remington}, B.~A., {Arnett}, D., {Drake}, R.~P., \& {Takabe}, H. 1999,
  Science, 284, 1488

\bibitem[{{Remington} {et~al.}(1997){Remington}, {Kane}, {Drake},
  {Glendinning}, {Estabrook}, {London}, {Castor}, {Wallace}, {Arnett}, {Liang},
  {McCray}, {Rubenchik}, \& {Fryxell}}]{1997PhPl....4.1994R}
{Remington}, B.~A., {Kane}, J., {Drake}, R.~P., {et~al.} 1997, Physics of
  Plasmas, 4, 1994

\bibitem[{{Robey} {et~al.}(2001){Robey}, {Kane}, {Remington}, {Drake},
  {Hurricane}, {Louis}, {Wallace}, {Knauer}, {Keiter}, {Arnett}, \&
  {Ryutov}}]{2001PhPl....8.2446R}
{Robey}, H.~F., {Kane}, J.~O., {Remington}, B.~A., {et~al.} 2001, Physics of
  Plasmas, 8, 2446

\bibitem[{{Schekochihin} {et~al.}(2004){Schekochihin}, {Cowley}, {Taylor},
  {Maron}, \& {McWilliams}}]{2004ApJ...612..276S}
{Schekochihin}, A.~A., {Cowley}, S.~C., {Taylor}, S.~F., {Maron}, J.~L., \&
  {McWilliams}, J.~C. 2004, \apj, 612, 276

\bibitem[{{Shakura} \& {Sunyaev}(1973)}]{1973AnA....24..337S}
{Shakura}, N.~I., \& {Sunyaev}, R.~A. 1973, \aap, 24, 337

\bibitem[{{Shvarts} {et~al.}(1995){Shvarts}, {Alon}, {Ofer}, {McCrory}, \&
  {Verdon}}]{1995PhPl....2.2465S}
{Shvarts}, D., {Alon}, U., {Ofer}, D., {McCrory}, R.~L., \& {Verdon}, C.~P.
  1995, Physics of Plasmas, 2, 2465

\bibitem[{{Smalyuk} {et~al.}(2005){Smalyuk}, {Sadot}, {Delettrez},
  {Meyerhofer}, {Regan}, \& {Sangster}}]{2005PhRvL..95u5001S}
{Smalyuk}, V.~A., {Sadot}, O., {Delettrez}, J.~A., {et~al.} 2005, Physical
  Review Letters, 95, 215001

\bibitem[{{Thornber} {et~al.}(2011){Thornber}, {Drikakis}, {Youngs}, \&
  {Williams}}]{2011PhFl...23i5107T}
{Thornber}, B., {Drikakis}, D., {Youngs}, D.~L., \& {Williams}, R.~J.~R. 2011,
  Physics of Fluids, 23, 095107

\bibitem[{{Truelove} \& {McKee}(1999)}]{1999ApJS..120..299T}
{Truelove}, J.~K., \& {McKee}, C.~F. 1999, \apjs, 120, 299

\bibitem[{{Truelove} \& {McKee}(2000)}]{2000ApJS..128..403T}
---. 2000, \apjs, 128, 403

\bibitem[{{Verdon} {et~al.}(1982){Verdon}, {McCrory}, {Morse}, {Baker},
  {Meiron}, \& {Orszag}}]{1982PhFl...25.1653V}
{Verdon}, C.~P., {McCrory}, R.~L., {Morse}, R.~L., {et~al.} 1982, Physics of
  Fluids, 25, 1653

\bibitem[{{Warren} {et~al.}(2005){Warren}, {Hughes}, \&
  {Badenes}}]{2005AAS...20717212W}
{Warren}, J.~S., {Hughes}, J.~P., \& {Badenes}, C. 2005, in Bulletin of the
  American Astronomical Society, Vol.~37, American Astronomical Society Meeting
  Abstracts, 172.12

\bibitem[{{Woosley} {et~al.}(2011){Woosley}, {Kerstein}, \&
  {Aspden}}]{2011ApJ...734...37W}
{Woosley}, S.~E., {Kerstein}, A.~R., \& {Aspden}, A.~J. 2011, \apj, 734, 37

\bibitem[{{Woosley} {et~al.}(2009){Woosley}, {Kerstein}, {Sankaran}, {Aspden},
  \& {R{\"o}pke}}]{2009ApJ...704..255W}
{Woosley}, S.~E., {Kerstein}, A.~R., {Sankaran}, V., {Aspden}, A.~J., \&
  {R{\"o}pke}, F.~K. 2009, \apj, 704, 255

\bibitem[{{Wunsch} \& {Kerstein}(2001)}]{2001PhFl...13..702W}
{Wunsch}, S., \& {Kerstein}, A. 2001, Physics of Fluids, 13, 702

\bibitem[{Youngs(1989)}]{youngs1989modelling}
Youngs, D.~L. 1989, Physica D: Nonlinear Phenomena, 37, 270

\bibitem[{{Zrake} \& {MacFadyen}(2013)}]{2013ApJ...769L..29Z}
{Zrake}, J., \& {MacFadyen}, A.~I. 2013, \apjl, 769, L29

\end{thebibliography}



\end{document}